\documentclass[twocolumn, pra, nofootinbib, aps]{revtex4}
\usepackage[utf8]{inputenc}

\usepackage{graphicx}
\usepackage{bm}
\usepackage{makeidx} 
\usepackage{amsmath}
\usepackage{amssymb}
\usepackage{underscore}
\usepackage{color}
\usepackage{amsthm}
\usepackage{hyperref}
\usepackage{cleveref}
\usepackage{diagbox}
\hypersetup{colorlinks, citecolor=magenta, filecolor=blue, linkcolor=blue, urlcolor=green}
\usepackage[normalem]{ulem}
\usepackage{float}
\usepackage{array}

\begin{document}

\title{Localizing genuine multimode entanglement: Asymmetric gains via non-Gaussianity
} 

\author{Ratul Banerjee\(^1\), Saptarshi Roy\(^1\), Tamoghna Das\(^{2}\), Aditi Sen(De)\(^1\)}

\affiliation{\(^1\)Harish-Chandra Research Institute, HBNI, Chhatnag Road, Jhunsi, Allahabad 211 019, India}
\affiliation{\(^2\)International Centre for Theory of Quantum Technologies,  University of Gda\'{n}sk, 80-952 Gda\'{n}sk, Poland.}

\begin{abstract}
Measurement-based quantum correlation  mimics several characteristics of multipartite quantum correlations and at the same time, it reduces the parent system to a smaller subsystem. On the other hand, genuine multipartite entanglement measures can capture certain features of a multisite composite system  that are inaccessible via bipartite quantum correlation quantifiers. We merge these two concepts by introducing localizable genuine multimode entanglement for continuous variable systems, both for Gaussian and non-Gaussian multimode parent states. We report a compact form of localizable generalized geometric measure for  multimode Gaussian states  when Gaussian measurements are performed in  some of the modes. We show that non-Gaussian measurements can concentrate more genuine multimode entanglement compared to the Gaussian ones. For non-Gaussian states with non-Gaussian measurements, we find that although four-mode squeezed vacuum state has permutation symmetry with respect to the exchange of first and third modes as well as the second and the fourth modes, the symmetry can be broken by performing measurements in  one of the modes in case of addition while for subtraction, such symmetry is preserved, thereby providing a   method for distinguishing multimode photon-added and -subtracted states via localizations. 
\end{abstract}

\maketitle

\section{Introduction}
\label{sec:intro}
The ``weirdness'' of quantum physics have puzzled scientists over the years although these peculiarities, absent in classical resources,  turn out to be boons for designing  quantum   information processing  tasks \cite{epr, Bellbook, sc1, sc2, mp, cglmp}. 
A conceptual revolution 
has paved the way for the discovery of remarkable protocols like quantum teleportation \cite{tele1,tele2,tele3}, dense coding \cite{dc,dcrev}, quantum key distribution (QKD) \cite{crypto1,crypto2}, quantum sensing \cite{qs}, quantum computation \cite{qcomp1, supcond2} and other quantum technologies like quantum memories \cite{qmem1, qmem2} and quantum batteries \cite{qb} to name a few. Most of the protocols, showing better efficiencies than their classical counterparts, employ some form of  quantum resources \cite{rt} which include  quantum coherence \cite{coherencerev}, entanglement \cite{hhhh}, quantum correlations, independent of entanglement \cite{ModiDiscord, discrevus}.   
Therefore, classification of nonclassical resources has utmost importance both from the perspectives of better understanding of quantum mechanics and utilizing the quantum fuel to build the quantum technologies.

We will be concentrating here on entanglement for which the resource theory is well-established in the literature, especially for  two spatially separated systems.  In this bipartite regime, 
 the categorization takes a simpler form and hence the detection as well as quantification are comparatively easier than the system comprising more than two parties.   
 However, multipartite entanglement is shown to be useful  for information processing tasks like in quantum networks and in measurement-based quantum computation \cite{g1, g2, onewayQC}. 
A multiparty pure state is said to be genuinely multipartite  entangled if  it is not product across every bipartition.
Based on the geometry of quantum states, the generalized geometric measure (GGM) \cite{ggm1} (see also \cite{ggm22,ggm3,ggm4,ggm5}) attempts to quantify the genuine multiparty entanglement  content of a state by computing the minimum distance of a given state from the set of non-genuinely multipartite entangled states (cf. \cite{ggm2, ggmmixed1}). 
On the other hand, multipartite entanglement content can also be effectively    characterized if  local measurements are performed on the subset of all the parties, thereby localizing entanglement in the unmeasured parties, referred to entanglement of assistance or  localizable entanglement \cite{divincenzo1998, smolin2005,  popp2005, gour2006, sadhukhan2017, amaro2018}. 

Given the resources, the next thing  is to consider the platform on which the quantum communication and computing tasks can be  performed. Several effective substrates have been identified that include cold atomic systems \cite{coldatom1, coldatom2, coldatom3, coldatom4, coldatom5, coldatom6, coldatom7}, superconducting qubits \cite{supcond, supcond2}, nuclear magnetic resonance (NMR) molecules \cite{nmr1,nmr2, nmr3, nmr4, nmr5, nmr6},  photonic systems \cite{photon1, photon2}  and many others. In this work, we investigate the classification of measurement-based genuine multimode entanglement, GME, in continuous variable states of light which can be realized in optical bosonic modes. 
In contrast to the issues faced when discrete polarization states of light  have been used for realizing several recent groundbreaking communication protocols, continuous  variable  (CV)  systems \cite{cv1, cv2} can overcome certain difficulties present in photonic qubits, like Bell-basis distinguishability. In particular,  CV states can  be  prepared with almost unit probability by using the nonlinear interaction of a crystal with a laser \cite{cv1}. All these advantages make CV systems a prominent physical system for building quantum gadgets.

Our aim is to calibrate entanglement present in the multimode CV systems via measurement process. Specifically, we perform both Gaussian as well as non-Gaussian local  measurements in one or many modes of a multimode Gaussian and non-Gaussian parent state to localize the GME of the whole state into the  unmeasured  modes consisting of multiple modes. The  entanglement generated in the  unmeasured modes is quantified by generalized geometric measure \cite{ggmcv}, leading to a localizable genuine multimode entanglement (LGME). In the CV sector, localization of bipartite entanglement for Gaussian states is discussed  \cite{firusek, adessorecent}. 
We systematize our investigations by splitting our analysis into three parts. First, we restrict ourselves to the Gaussian regime where we examine LGME of a Gaussian state, namely a four-mode squeezed vacuum (FMSV) state \cite{fmsv}, by employing optimal Gaussian measurements in one of the four modes of the FMSV. 
Secondly, we  replace the Gaussian measurements with the non-Gaussian ones that involve photon counting schemes. Note that the chosen non-Gaussian measurement scheme which can be easily performed experimentally, although there can be other optimal ones. 
However, our analysis reveals that LGME obtained from non-Gaussian measurements is always higher than the one obtained with the optimal Gaussian measurements.  

Finally, we take up the most general case where both the states and measurements are non-Gaussian. A prominent and efficient method to create non-Gaussian states is to add (subtract) photons in Gaussian states. Moreover, it was shown that non-Gaussianity, both in the form of states and operations,  can be useful for several quantum information processing tasks like distillation, error corrections, quantum sensing than  Gaussian states and operations \cite{qcomp, bc1, Ent-distillation, Giedke-ent-distillation, qcloning, gaussian-error, qmetro, phase-estimation, qcommunication, entdistribution}.  
By employing the same non-Gaussian measurements, we report  that  photon subtracted  FMSV can localize more GME than that of the photon-added state.  In this respect, we first notice that four-mode Gaussian states is symmetric with the permutation of first and third as well as second and fourth modes.    We  then show that if one performs measurement in the fourth mode, the addition of photons in the second mode which is unmeasured leads to a higher multimode entanglement than that of the fourth mode, thereby showing symmetry-breaking nature induced by measurements. Interestingly, in case of subtraction, measurement is unable to break the permutation symmetry present in the four-mode non-Gaussian states.  Therefore, LGME can act as a method to discriminate multimode photon-subtracted states  from the photon-added ones. Moreover,  such contrasting behavior can  furnish a mechanism by which the modes used for photon addition can be detected. 

The paper is organized as follows. In Sec. \ref{sec:introGME}, we introduce localizable genuine multimode entanglement or measurement-based multimode entanglement. After setting the stage, in Sec. \ref{sec:gaussian}, we derive a compact form of LGME for Gaussian states with optimal Gaussian measurement and also show that non-Gaussian measurements can perform better than the optimal Gaussian measurement in the process of localization. In Sec. \ref{sec:nongaussian}, we investigate LGME in the photon-added and -subtracted states. 
We finally conclude in Sec. \ref{sec:con}.


\section{Localizable Genuine Multimode  Entanglement}
\label{sec:introGME}

Let us  introduce the localizable genuine multimode entanglement for an arbitrary  $N$-mode state. To localize entanglement, local Gaussian or non-Gaussian measurements are performed on $L$ modes 
 of an $N$-mode state. Since we are interested in the residual multimode entanglement,  we must have $N - L=k \geq 3$. 
For a given multimode entanglement measure, $\mathcal{E}$, the localizable or measurement-based genuine multimode entanglement can then be defined as 
\begin{eqnarray}
\mathcal{L}\mathcal{E} = \max_{\{\mathcal{M}\}} \int p_k \mathcal{E}(|\psi\rangle_k),
\label{eq:LGME}
\end{eqnarray}
 where maximization is performed over the local measurements on $L$ modes, and the averaging is performed over  the measurement outcomes of the resulting state of $k \geq 3$-modes. 
 
In this paper, we take an arbitrary four-mode squeezed vacuum state as the parent state, and by performing a measurement on a single mode,  we obtain the resulting three-mode state. Depending on the types of measurement and the initial state, the resulting state is either Gaussian or non-Gaussian in nature. After localization,  multimode entanglement  is quantified  geometrically, i.e. via generalized geometric measure \cite{ggm1, ggm22, ggm3,   ggm4, ggm5, ggm2, ggmmixed1, ggmcv}.  For a $N$-mode state, $|\psi_N\rangle$, it is defined   via Fubini study metric \cite{fubini1,fubini2} as 
\begin{eqnarray}
\label{Eq:GGM}
{\cal G}(|\psi_{N}\rangle) = 1 - \max_{|\chi\rangle \in nG} |\langle \chi |\psi_{N}\rangle|^2 = 1 -  \nonumber \\
\max \big \lbrace \lambda_{\cal A:B} | {\cal A}\cup {\cal B} = \lbrace 1,2,\ldots, N \rbrace, {\cal A}\cap{\cal B} = \emptyset \big \rbrace.
\end{eqnarray}
The maximization is taken over the set  of  $N$-mode pure states, $\{|\chi\rangle\}$,  which are not  genuinely multimode entangled and the set is denoted by \(nG\). Interestingly, 
the optimization can be conquered by using  Schmidt decomposition for continuous variable systems \cite{schmidt,schmidt2} which can be easily extended for any normalizable infinite dimensional state. For four-mode Gaussian states, the details will be presented  in the succeeding  section. 


\section{Gaussian genuine Localizable  entanglement: Gaussian vs. Non-Gaussian measurements}
\label{sec:gaussian}

Before computing LGME for four-mode squeezed vacuum state, let us first briefly describe the method to compute GGM for multimode state. 
Upto local displacement operations, a  Gaussian state is specified uniquely by its covariance matrix \cite{photon1, photon2}. 
For an arbitrary $m$-mode Gaussian state, $\rho$, the covariance matrix, $\Lambda$,  is a $2m \times 2m$ matrix,  defined as
$\Lambda_{ij} = \frac{1}{2} \big \langle \lbrace R_i,R_j \rbrace \big \rangle - \langle R_i  \rangle \langle R_j \rangle$,
where $\vec{R} = (q_1,p_1,q_2,p_2, ... q_m,p_m)^{\text{T}}$. Here $q_i$s and $p_i$s are the usual phase space quadrature operators which can be
given  in terms of raising and lowering operators as
$q_j = \frac{1}{\sqrt{2}}(a_j + a_j^{\dagger}), ~p_j = \frac{1}{\sqrt{2}i}(a_j - a_j^{\dagger})$,
where $i = \sqrt{-1}$. 
 If  $\rho$  has to be a valid density matrix,  the covariance matrix in terms of the symplectic matrix, $J$, has to satisfy
$\Lambda+iJ \geq 0, \text{ where } J = \bigoplus_{i=1}^{m} 
\begin{bmatrix}
 0 & 1 \\
-1 & 0 
\end{bmatrix}$. 
By using  Williamson's theorem \cite{contvar-rev,ferraro}, the covariance matrix $\Lambda$ can be obtained from $\Lambda^d$ by appropriate symplectic transformation ($\textbf{S}_\Lambda$),
$\Lambda = \textbf{S}_\Lambda \Lambda^d \textbf{S}_\Lambda^{\text{T}}$
 with $\Lambda^d = \bigoplus_{i=1}^m \nu_i \mathbb{I}_2$,
where $\{\nu_i\}$s are the symplectic eigenvalues of $\Lambda$ and $\mathbb{I}_2$ is the $2 \times 2$ identity matrix.  
The GGM ($\mathcal{G}$) of a $N$-mode pure Gaussian state $|\psi_{N}\rangle$  in terms of symplectic eigenvalues takes the compact form as \cite{ggmcv}
\begin{eqnarray}
\label{eq:GGM-new-form}
\mathcal{G}(|\psi_{N}\rangle) = 1 - \max \mathcal{P}_m \Big\lbrace \prod_{i=1}^m \frac{2}{1 +2 \nu_i} \Big\rbrace_{m=1}^{\big[\frac{N}{2}\big]}. 
\end{eqnarray}
Here maximization is performed by considering symplectic eigenvalues of all the reduced states of $|\psi\rangle_{N}$ with $m$-modes which are obtained by applying the permutation  operator, $\mathcal{P}_m$ while $[x]$ denotes the integral part of $x$.

\subsection{Concentrating genuine entanglement of Gaussian states with  Gaussian measurement}

To compute LGME with the Gaussian state, we consider the four-mode squeezed vacuum  state \cite{fmsv} as the parent state on which Gaussian measurements are performed.  The FMSV state  with squeezing strength $r$  can be prepared in the laboratories  by using linear optical elements like $50$:$50$ beam splitters and two single-mode squeezed vacuum states and   can be written as
\begin{eqnarray}
\label{eq:FMSV}
&& |FMSV\rangle \nonumber \\
&=& e^{\frac r2 \sum_{i=1}^4a_i^\dagger a_{i+1}^\dagger - a_i a_{i+1}}|0000\rangle  \nonumber \\
&=& \frac{1}{\cosh r} \sum_{n = 0}^\infty \left(\frac 12 \tanh r\right)^n \sum_{r_1 = 0}^n \sum_{r_2 = 0}^n 
\sqrt{\binom{n}{r_1}} \sqrt{\binom{n}{r_2}} \nonumber \\
&&  \hspace{1.3in} |n - r_1\rangle |n - r_2\rangle | r_1\rangle | r_2\rangle,
\end{eqnarray}
Here for $i=4$, $i+1$ is considered to be $1$.
The GGM in this case  reads as 
\begin{eqnarray}
\label{eq:GGMFMSV}
\mathcal{G}(|FMSV\rangle) = 1 - \max \Big\lbrace \frac{2}{1+\cosh^2 r}, \nonumber \\ \frac{2}{1+\cosh 2r}, \Big(\frac{2}{1+\cosh r}\Big)^2 \Big\rbrace. 
\end{eqnarray}
It is important to note  here that the FMSV state has an inherent permutation symmetry -- it remains invariant under the exchange of first and third mode as well as the second and fourth mode.

\begin{figure}
\includegraphics[width=\linewidth]{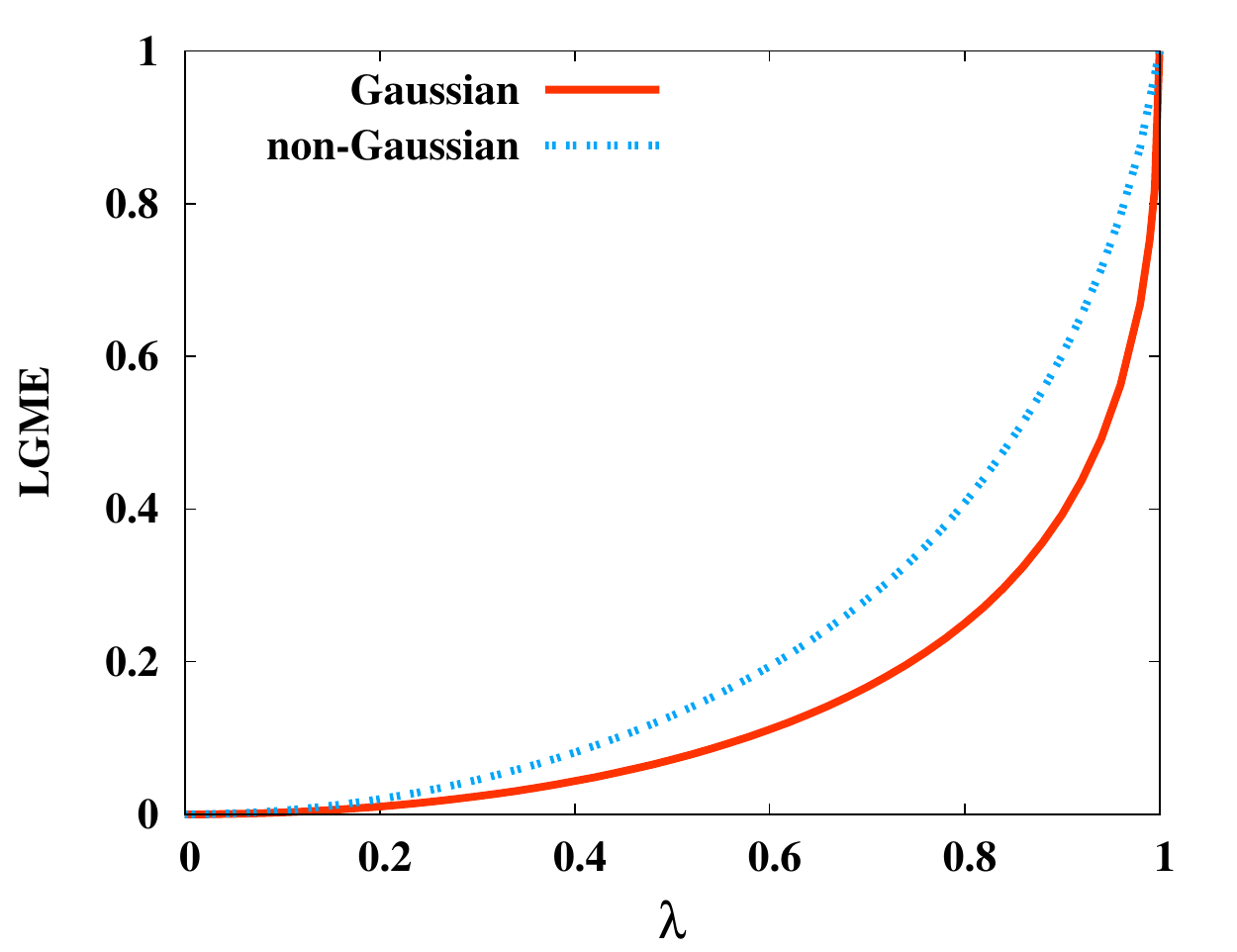}
\caption{Gaussian vs. non-Gaussian measurements. Taking four-mode squeezed vacuum state, given in Eq. (\ref{eq:FMSV}), as the parent state, genuine multimode entanglement (ordinate) is computed with respect to the squeezing parameter, $\lambda = \tanh r$ (abscissa). The measurement is performed in the fourth mode. The solid line represents the LGME of the FMSV state after optimal  Gaussian measurements while the dashed line is for the same with non-Gaussian measurements. Clearly, photon counting non-Gaussian measurements are advantageous to localize multimode entanglement compared to the optimal Gaussian ones. Both the axes are dimensionless.     }
\label{fig:GvsnG}
\end{figure}

Without loss of generality, let us perform a projective Gaussian measurement in the fourth mode of the FMSV state. The most general covariance matrix of a single-mode Gaussian state with which the measurement is performed is the   squeezed coherent state whose covariance matrix can be represented as  
\small
\begin{eqnarray}
 && \sigma_m = \nonumber \\ 
 &&\begin{bmatrix}
 \cosh{2r'} +\cos{\phi}\sinh{2r'} - &\sin{\phi}\sinh{2r'} \\ 
 \sin{\phi}\sinh{2r'} & \cosh{2r'}- \cos{\phi}\sinh{2r'}
\end{bmatrix}, \nonumber \\ 
\end{eqnarray}\\
\normalsize
with \(r'\) being the squeezing parameter, and \(\phi\) being the angle of the quadrature. 
It turns out that the reduced covariance matrices for the various measurement outcomes are identical while only the displacement vectors are outcome-dependent. Since the GGM content depends only on the covariance matrix, we do not require to perform averaging  over the different measurement outcomes involved in the definition (Eq. (\ref{eq:LGME})). The post-measured three-mode covariance matrix, $(\Lambda_m)$, can be computed as 
 \begin{equation}
  \Lambda_{m}= \sigma_{123} - \sigma_{c}.((\sigma_{m} +\sigma_{4})^{-1}).\sigma_{c}^{T},
\end{equation}
where $\sigma_{123}$ and $\sigma_4$ denote the first three modes and the fourth mode respectively of the covariance matrix of  \(|FMSV\rangle\), $\Lambda^{\text{FMSV}}$, which is given by 
\begin{eqnarray}
\frac{1}{2}
\begin{bmatrix}
\cosh^2 r ~\mathbb{I}_2 & \frac{1}{2}\sinh 2r ~\sigma_z & \sinh^2 r ~\mathbb{I}_2 & \frac{1}{2}\sinh 2r ~\sigma_z \\
\frac{1}{2}\sinh 2r ~\sigma_z & \cosh^2 r ~\mathbb{I}_2 & \frac{1}{2}\sinh 2r ~\sigma_z & \sinh^2 r ~\mathbb{I}_2 \\
\sinh^2 r ~\mathbb{I}_2 & \frac{1}{2}\sinh 2r ~\sigma_z & \cosh^2 r ~\mathbb{I}_2 & \frac{1}{2}\sinh 2r ~\sigma_z \\
\frac{1}{2}\sinh 2r ~\sigma_z & \sinh^2 r ~\mathbb{I}_2 & \frac{1}{2}\sinh 2r ~\sigma_z & \cosh^2 r ~\mathbb{I}_2 \nonumber 
\end{bmatrix},
\end{eqnarray}
with \(\mathbb{I}_2\) and \(\sigma_z\) being the identity and Pauli matrix in the z-direction respectively.  
 Here,   $\sigma_c$ represents the correlation matrix between the first three and the fourth mode of  FMSV. In particular, they are given by
\begin{eqnarray}
\sigma_{123} = \frac{1}{2}
\begin{bmatrix}
\cosh^2 r ~\mathbb{I}_2 & \frac{1}{2}\sinh 2r ~\sigma_z & \sinh^2 r ~\mathbb{I}_2  \\
\frac{1}{2}\sinh 2r ~\sigma_z & \cosh^2 r ~\mathbb{I}_2 & \frac{1}{2}\sinh 2r ~\sigma_z  \\
\sinh^2 r ~\mathbb{I}_2 & \frac{1}{2}\sinh 2r ~\sigma_z & \cosh^2 r ~\mathbb{I}_2 \nonumber 
\end{bmatrix},
\end{eqnarray}
\begin{eqnarray}
\sigma_4 = \frac12 \cosh^2 r ~\mathbb{I}_2, ~ \sigma_c = \frac12 \begin{bmatrix}
 \frac{1}{2}\sinh 2r ~\sigma_z \\
 \sinh^2 r ~\mathbb{I}_2 \\
 \frac{1}{2}\sinh 2r ~\sigma_z  \nonumber 
\end{bmatrix}.
\end{eqnarray}
Note that $\Lambda_m \equiv \Lambda_m(r,r',\phi)$, and the corresponding localizable GME in terms of GGM can be expressed as
\begin{eqnarray}
\mathcal{L}\mathcal{G}(\Lambda^{\text{FMSV}}(r)) = \max_{r',\phi} \mathcal{G}(\Lambda_m(r,r',\phi)).
\end{eqnarray}
Our analysis reveals that the optimal measurement configuration turns out to be homodyne detection of either $x$ or $p$ quadrature. Hence, for our analysis, without loss of  generality,  we consider  the $x$-quadrature. Therefore, the maximal LGME is obtained for $r' \rightarrow \infty$ and $\phi = 0$.  In this optimal configuration, $\Lambda_m^{opt}$ takes a simple form and its three single mode reduced covariance matrices, $R_1, R_2$, and $R_3$, are  given by
\begin{eqnarray}
R_1 = R_3 = \frac12
\begin{bmatrix}
\cosh^2 r & 0 \\
0 & 1 
\end{bmatrix},
\label{eq:E1}
\end{eqnarray}
and
\begin{eqnarray}
R_2  = \frac12
\begin{bmatrix}
\cosh^2 r & 0 \\
0 & 1+ \tanh^2 r 
\end{bmatrix}.
\label{eq:E2}
\end{eqnarray}
The symplectic eigenvalues of $R_{1(3)}$ and $R_2$ are computed to be $\nu_1 = \frac12 \cosh r$ and $\nu_2 = \frac12 \sqrt{\cosh 2r}$ which leads to the LGME of the FMSV state using Gaussian measurements as 
\begin{eqnarray*}
\mathcal{L}\mathcal{G}(\Lambda^{\text{FMSV}}) &=&  \mathcal{G}(\Lambda_m^{opt}) \nonumber \\
&=&1 - \max \Big \lbrace \frac{2}{1 + \cosh r}, \frac{2}{1 + \sqrt{\cosh 2r}} \Big \rbrace. 
\end{eqnarray*} 
However, we find that for all $r>0$, $\cosh r < \sqrt{\cosh 2r}$, and, therefore, $\frac{2}{1 + \cosh r}$ is always larger, thereby providing a compact form of LGME for FMSV state with optimal local Gaussian measurement as
\begin{eqnarray}
\mathcal{L}\mathcal{G}(\Lambda^{\text{FMSV}}) =  \tanh^2 \frac{r}{2}.
\end{eqnarray} 
Clearly, it increases monotonically with the variation of the squeezing parameter, $r$ as shown in Fig. \ref{fig:GvsnG}.

\subsection{Non-Gaussian measurement enhances localization}
\label{sec:G-nG}

Let us now see whether more entanglement can be localized from a Gaussian state, when instead of optimal Gaussian measurements, one employs some non-Gaussian measurement schemes. As  non-Gaussian measurements,  the photon counting operations are performed in any one of the modes of the Gaussian state and  can be translated  to a measurement in the Fock basis $\lbrace |k\rangle \rbrace$, where $|k\rangle$ denotes the \(k\)-photon state. When the fourth mode of $|FMSV\rangle$ is measured in the Fock basis and  $|k\rangle$ clicks, the resulting normalized three-mode post-measured state is denoted by $|\psi_k\rangle$, where we have $\langle k|FMSV \rangle = \sqrt{p_k}|\psi_k\rangle$ with $p_k$ being the clicking probability. 

The LGME content of $|FMSV\rangle$ under the photon counting (non-Gaussian) measurement procedure reads as
\begin{eqnarray}
\label{eq:LGGMnonG}
\mathcal{L}\mathcal{G}(|FMSV\rangle) = \sum_k p_k \mathcal{G}(|\psi_k\rangle).
\end{eqnarray} 
In Fig. \ref{fig:GvsnG},  we observe that by using the photon counting non-Gaussian measurement, higher amount of GME  can be localized than the obtained via  the optimal Gaussian measurement. Notice for three-mode Gaussian state, such an advantageous scenario with non-Gaussian measurement was also  reported \cite{firusek}. 

\begin{figure*}
\includegraphics[width=\linewidth]{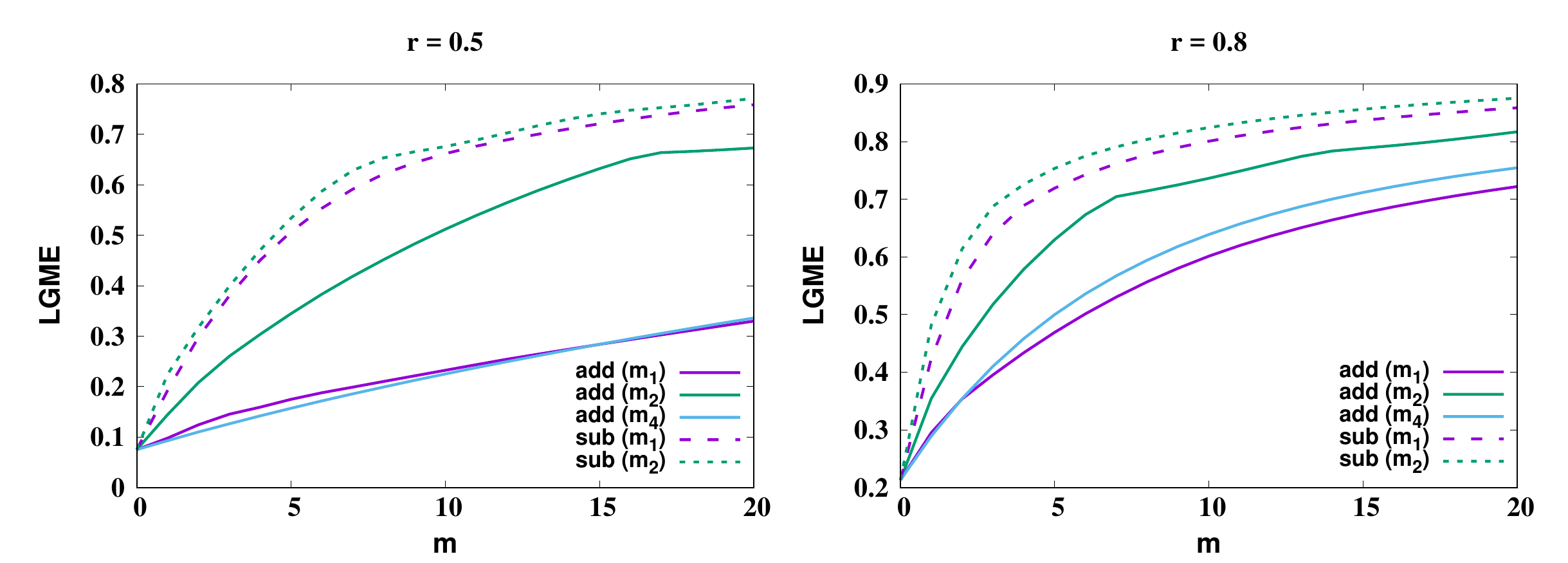}
\caption{ Localizable genuine multimode entanglement (vertical axis) against  the number of photons added and subtracted (horizontal axis), $m$, from a single mode of the FMSV state. The solid line is for adding photons from a single mode while the dashed line represents  subtraction of photons from the same. In legend,  the subscript, \(i = 1, 2, 4\)s in \(m_i\) indicate the mode in which photons are added (subtracted). First of all, localizable entanglement gets enhanced due to the addition (subtraction) of photons. 
We  observe that the LGME for adding or subtracting photons in (from) mode \(1\) or  \(3\)  as well as subtracting photons in mode \(2\) or \(4\) give rise to the same value while adding photons in the second and fourth mode leads to a different entanglement content. Such an observation is interesting since the parent state remains same with the interchange of mode \(1\) and \(3\) as well as mode \(2\) and \(4\).    Both the axes are dimensionless.    }
\label{fig:LGMEaddsubonemode}
\end{figure*}

\section{ more localization of Genuine multimode  entanglement for photon-subtracted states}
\label{sec:nongaussian}

In this section, we investigate the role of non-Gaussianity, both at the level of states and measurements for concentrating entanglement into smaller subsystems. Our analysis in the previous section already hinted at the enhancement properties of LGME with non-Gaussian measurements. Here we examine the most general case of localization of GME for non-Gaussian states by employing non-Gaussian measurements.
We have already discussed the non-Gaussian measurement strategy in the previous section. Now we discuss the procedure for deGaussification of the FMSV state.

Among the various methods by which a state can be deGaussified, we choose to employ photon addition and subtraction operations. Such a choice is based on two important features  -- these photonic operations can be implemented experimentally, and secondly they have been shown to have enhancing effects in varied scenarios ranging from bipartite entanglement,  violation of Bell inequality, to genuine multimode entanglement in the continuous variable sector, thereby showing its possible applications in quantum information processing tasks \cite{TMSV, enhance1, enhance2, enhance3, bell, exp1, exp2, ggmcv, fmsv}.  LGME, in this case, will be denoted as \(\mathcal{LG}^{add}\) and \(\mathcal{LG}^{subtract}\) to represent photon-addition and -subtraction respectively. 

The photon-added and -subtracted FMSV states with $m_i$ number of photons are added or subtracted in or from the mode $i$ $(i= 1, 2, 3, 4)$ respectively read as 
\begin{widetext}
\begin{eqnarray}
|\psi^{add\{m_i\}}_{FMSV}\rangle = \frac{1}{\sqrt{N^{add}}}  \sum_{n = 0}^\infty \sum_{r_1 = 0}^n \sum_{r_2 = 0}^n \left(\frac 12 \tanh r\right)^n \sqrt{\binom{n}{r_1}} \sqrt{\binom{n}{r_2}}  \sqrt{\frac{(n -r_1 + m_1)!}{(n - r_1)!}}  \sqrt{\frac{(n -r_2 + m_2)!}{(n - r_2)!}}  \nonumber \\
\sqrt{\frac{(r_1 + m_3)!}{r_1!}} \sqrt{\frac{(r_2 + m_4)!}{r_2!}} |n - r_1 + m_1\rangle_1 |n - r_2 + m_2\rangle_2 | r_1 + m_3\rangle_3 | r_2 + m_4\rangle_4, \label{eq:add-FMSV}\\
|\psi^{sub\{m_i\}}_{FMSV}\rangle = \frac{1}{\sqrt{N^{sub}}}  \sum_{n = M}^\infty \sum_{r_1 = m_3}^{n - m_1} \sum_{r_2 = m_4}^{n - m_2} \left(\frac 12 \tanh r\right)^n \sqrt{\binom{n}{r_1}} \sqrt{\binom{n}{r_2}}  \sqrt{\frac{(n -r_1 )!}{(n - r_1 - m_1)!}}  \sqrt{\frac{(n -r_2 )!}{(n - r_2 - m_2)!}}  \nonumber \\
\sqrt{\frac{r_1!}{(r_1 - m_3)!}} \sqrt{\frac{r_2!}{(r_2 - m_4)!}} |n - r_1 - m_1\rangle_1 |n - r_2 - m_2\rangle_2 | r_1 - m_3\rangle_3 | r_2 - m_4\rangle_4, \label{eq:sub-FMSV}
\end{eqnarray}
\end{widetext}
where $M = \max [m_1 + m_3, m_2 + m_4]$ and, $N^{add}$ and $N^{sub}$ are the respective normalization constants and can be expressed as follows
\begin{eqnarray}
N^{add} =  \sum_{n = 0}^\infty \left(\frac 12 \tanh r\right)^{2n} \sum_{r_1,r_2 = 0}^n   {\binom{n}{r_1}} {\binom{n}{r_2}} \nonumber \\
 \frac{(n -r_1 + m_1)!}{(n - r_1)!}  \frac{(n -r_2 + m_2)!}{(n - r_2)!} \frac{(r_1 + m_3)!}{r_1!} \frac{(r_2 + m_4)!}{r_2!}, ~~~
 \end{eqnarray}
 \begin{eqnarray}
 N^{sub} =  \sum_{n = M}^\infty \left(\frac 12 \tanh r\right)^{2n} \sum_{r_1 = m_3}^{n - m_1 } {\binom{n}{r_1}} \sum_{r_2 = m_4}^{n - m_2}    {\binom{n}{r_2} } \nonumber \\
 \frac{(n-r_1)!}{(n -r_1 - m_1)!} 
 \frac{(n - r_2)!}{(n - r_2 - m_2)!} \frac{r_1 !}{(r_1 - m_3)!} \frac{r_2!}{(r_2 - m_4)!}. ~~~
 \end{eqnarray}

Let us  first concentrate on localizing GME in $|\psi^{sub\{m_i\}}_{FMSV}\rangle$ via photon counting measurements in the fourth mode. When the $k$-photon state clicks in the fourth mode, the normalized post-measurement state is computed as
\begin{eqnarray}
&& |\psi^{sub\{m_i\}}_{FMSV}(k)\rangle = \frac{1}{\sqrt{p_k^{sub}}} {}_4\langle k |\psi^{sub\{m_i\}}_{FMSV}\rangle, \nonumber \\
&& = \sum_{n = n_{min}}^\infty \sum_{r_1 = 0}^{n + M - m_1 - m_3} f^{sub}_k(n,r_1, \{m_i\})  \nonumber \\ 
&&|n + M  - m_1  - m_3 -r_1 \rangle_1 |n + M - k - m_4 - m_2\rangle_2 | r_1 \rangle_3 ~~~~~~
\label{eq:pms-sub}
\end{eqnarray}
where $n_{min} = \max\, [0,k + m_2 + m_4 - M]$, and
\begin{eqnarray}
&& f^{sub}_k(n,r_1, \{m_i\}) \nonumber \\
&& = \frac{1}{\sqrt{p_k^{sub} N^{sub}}} \left(\frac 12 \tanh r\right)^n (n + M)! \frac{1}{\sqrt{r_1! k!}}\times \nonumber \\
&& \frac{1}{\sqrt{(n + M - r_1 - m_1 - m_3 )! }} \frac{1}{\sqrt{(n + M - k - m_2 - m_4 )! }}, \nonumber \\
\label{eq:fsub}
\end{eqnarray}
and the subscript in the first line of Eq. (\ref{eq:pms-sub}) indicates that the measurement is on the fourth mode. 
It occurs with probability,
\begin{eqnarray}
&& p_k^{sub} = |{}_4\langle k |\psi^{sub\{m_i\}}_{FMSV}\rangle|^2 \nonumber  \\ 
&& = \frac{1}{N^{sub}}\sum_{n = n_{min}}^\infty \left(\frac 12 \tanh r\right)^{2n} \sum_{r_1 = 0}^{n + M - m_1 - m_3} \frac{1}{r_1!} \frac{1}{k!} \nonumber \\  
&& \frac{(n +M)!}{(n +M - r_1 - m_1  - m_3)!}  \frac{(n + M)!}{(n + M - k - m_2 - m_4)!}.  ~~~~~~
\label{eq:pksub}
\end{eqnarray} 
The LGME content of $|\psi^{sub\{m_i\}}_{FMSV}\rangle$  can then be obtained by using Eq. (\ref{eq:LGGMnonG}) and 
the form of the photon-subtracted state leads to the following proposition.

\begin{figure}[t]
\includegraphics[width=\linewidth]{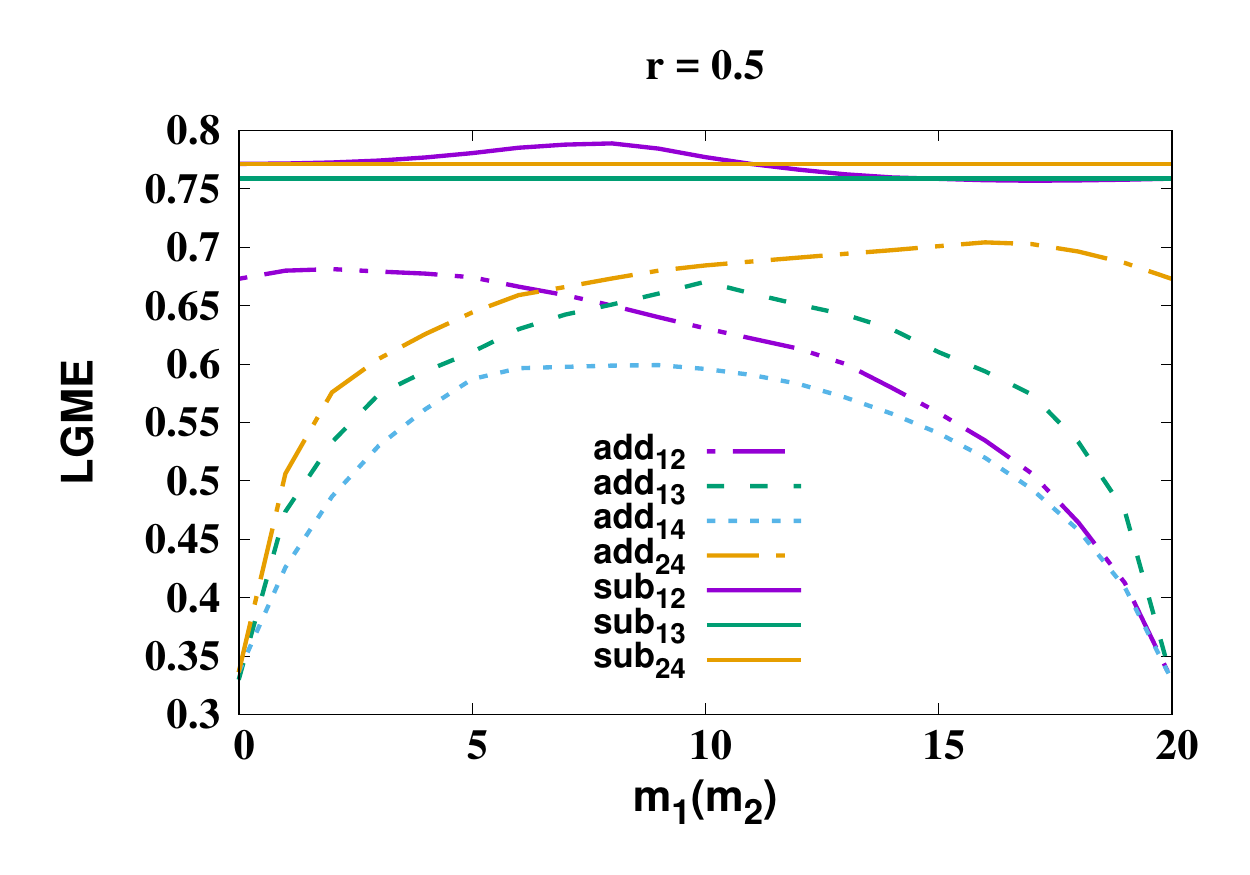}
\caption{ Localizable genuine multimode entanglement  (ordinate) vs.  the number of photons added to and subtracted from two different modes of the FMSV state (abscissa), where the total number of photons added or subtracted are fixed to $20$. Here \(m_i + m_j = 20\), where \(i, j = 1, 2, 3, 4\). . The label of the horizontal axis is for \(m_i\) where \(i<j\).   The dashed lines represent photon-addition while the solid line is for subtracting photons from the same. The legend, add\(_{ij}\) or sub\(_{ij}\) represent addition of photons in  \((i, j)\)-pair  of modes or subtraction of photons from the same.   When photons are subtracted from first and third modes or second and fourth modes, LGME remains constant  (see text for the proof). Both the axes are dimensionless. }
\label{fig:LGMEconatraint}
\end{figure}

\noindent\textbf{Proposition 1.} The localizable GME of the photon-subtracted state is a function of \(m_1 + m_3\) and \(m_2+ m_4\).

\begin{proof}
First we note that when the \(k\)th outcome appears after the photon-counting measurement of the fourth mode, the Schmidt coefficients of each resulting three-mode state  are functions of \(m_1 + m_3\) and \(m_2+ m_4\),  as seen in Eqs.  \eqref{eq:pms-sub}, \eqref{eq:fsub}, and \eqref{eq:pksub}.  And the GME in Eqs. (\ref{Eq:GGM}) and (\ref{eq:LGGMnonG})  is a function of Schmidt coefficients. Hence the proof.  \end{proof}

The two immediate corollaries of the above proposition are following:

\noindent\textbf{Corollary 1.} The localizable GME of the photon-subtracted state is invariant under the interchange of $m_1 \leftrightarrow m_3$ and $m_2 \leftrightarrow m_4$. 

\noindent\textbf{Corollary 2.} The localizable GME of the photon-subtracted state remains 
constant when photons are subtracted from the first and the third modes in a constraint manner, i.e., when $m_1 + m_3 = n$. Similarly for $m_2 + m_4 = n'$.

The post-measurement state of $|\psi^{add\{m_i\}}_{FMSV}\rangle$, when $k$th photon is detected in the fourth mode, is given by
\begin{eqnarray}
&& |\psi^{add\{m_i\}}_{FMSV}(k)\rangle = \frac{1}{\sqrt{p_k^{add}}} {}_4\langle k |\psi^{add\{m_i\}}_{FMSV}\rangle, \nonumber \\
&& = \sum_{n = k - m_4}^\infty \sum_{r_1 = 0}^{n } f^{add}_k(n,r_1, \{m_i\})  \nonumber \\ 
&& \hspace{0.5in}|n + m_1 -r_1 \rangle_1 |n - k + m_2 + m_4\rangle_2 | r_1 + m_3 \rangle_3, ~~~~~~
\label{eq:pms-add}
\end{eqnarray}
where 
\begin{eqnarray}
&& f^{add}_k(n,r_1, \{m_i\}) \nonumber \\
&& \hspace{-0.2in}= \frac{1}{\sqrt{p_k^{add} N^{add}}} \left(\frac 12 \tanh r\right)^n \sqrt{\binom{n}{r_1}} \sqrt{\binom{n}{k}} \sqrt{\frac{(n + m_1 - r_1)!}{(n-r_1)!}} \nonumber \\
&& \sqrt{\frac{(n-k+m_2+m_4)!}{(n-k+m_4)!}} \sqrt{\frac{(r_1 + m_3)!}{r_1!}} \sqrt{\frac{k!}{(k-m_4)!}}, ~~~
\label{eq:fadd}
\end{eqnarray}
and the probability of getting $k$th photon in the fourth mode is given by 
\begin{eqnarray}
&& p_k^{add} = |{}_4\langle k |\psi^{add\{m_i\}}_{FMSV}\rangle|^2 \nonumber  \\ 
&& \hspace{-0.2in}= \frac{1}{N^{add}}\sum_{n = k - m_4}^\infty \left(\frac 12 \tanh r\right)^{2n} \sum_{r_1 = 0}^{n }  {\binom{n}{r_1}} {\binom{n}{k}} \frac{(n + m_1 - r_1)!}{(n-r_1)!}\nonumber \\  
&& \frac{(n-k+m_2+m_4)!}{(n-k+m_4)!} \frac{(r_1 + m_3)!}{r_1!}\frac{k!}{(k-m_4)!}.  
\end{eqnarray} 
The form of 
 $|\psi^{add\{m_i\}}_{FMSV}\rangle$ leads to the computation of \(\mathcal{LG}^{add}\) which also possess the following symmetry:\\
\noindent\textbf{Proposition 2.} The localizable GME of the photon-added state is invariant under the interchange of the number of photons added in mode one and three, i.e., $m_1 \leftrightarrow m_3$.
\begin{proof}
When $m_i$  number of photons are added in mode $i$,  the corresponding state given in  Eq. (\ref{eq:pms-add}) can be expressed as 
\begin{eqnarray}
&& |\psi^{add\{m_i\}}_{FMSV}(k)\rangle \nonumber \\
&& = \sum_{n = k - m_4}^\infty \sum_{r_1 = 0}^{n } f^{add}_k(n,n - r_1, \{m_i\})  \nonumber \\ 
&& \hspace{0.5in}|r_1+ m_1  \rangle_1 |n - k + m_2 + m_4\rangle_2 | n - r_1 + m_3 \rangle_3, ~~~~~~
\end{eqnarray}
where we  replace $r_1 \rightarrow n - r_1$, and $f^{add}_k(n,n - r_1, \{m_i\})$ is given in Eq. (\ref{eq:fadd}). Now if instead of adding $m_1$ number of photons in the first mode, we add $m_3$ in mode one and $m_1$ in mode three, i.e., interchanging $m_1 \leftrightarrow m_3$, while $m_2$ and $m_4$ remain same, we have 
\begin{eqnarray}
&& |\psi^{add\{m_i\}}_{FMSV}(k)\rangle_{m_1 \leftrightarrow m_3} \nonumber \\
&& \hspace{-0.2in} = \sum_{n = k - m_4}^\infty \sum_{r_1 = 0}^{n } f^{add}_k (n,n - r_1, \{m_i\})_{m_1 \leftrightarrow m_3}  \nonumber \\ 
&& \hspace{0.2in}|r_1+ m_3  \rangle_1 |n - k + m_2 + m_4\rangle_2 | n - r_1 + m_1 \rangle_3, \nonumber \\
&& \hspace{-0.2in} = \sum_{n = k - m_4}^\infty \sum_{r_1 = 0}^{n } f^{add}_k(n,r_1, \{m_i\})  \nonumber \\ 
&& \hspace{0.2in}|r_1+ m_3  \rangle_1 |n - k + m_2 + m_4\rangle_2 | n - r_1 + m_1 \rangle_3. ~~~~~
\label{eq:pmsaddinterchange}
\end{eqnarray}
Hence, one can  easily see that Eq. (\ref{eq:pms-add}) is same as Eq. (\ref{eq:pmsaddinterchange}) with mode one and three being interchanged, which shows that the LGME is invariant under the interchange of $m_1 \leftrightarrow m_3$ in case of photon-addition.
\end{proof}


\noindent\emph{Single-mode operations.} The observations from the single-mode photon addition or subtraction can be listed as follows (see Fig. \ref{fig:LGMEaddsubonemode}):

\begin{enumerate}
    \item As obtained in case of other multimode entanglement measures \cite{TMSV, enhance1, enhance2, enhance3, bell},  we also observe here that the amount of entanglement localized in  modes gets increased 
    due to the addition (subtraction) of photons in the multimode squeezed vacuum states, irrespective of choice of the modes on which measurements are performed. 
    
    \item Unlike two-mode entanglement \cite{TMSV}, LGME is higher in the photon-subtracted state than that of the photon-added state (cf. \cite{ggmcv, fmsv}).  Note that the enhancement features of LGME on addition and subtraction of photons can be explained by the non-Gaussianity \cite{ng1, ng2, ng3} induced in the FMSV state by the photonic operations. However, non-Gaussianity cannot conclusively answer the question that which type of deGaussification (photon addition or subtraction) induces greater enhancement of LGME.

    \item As mentioned before, the four-mode squeezed vacuum state has a symmetry in the exchange of modes, i.e. \(1 \leftrightarrow 3\) and \(2 \leftrightarrow 4\) and hence its entanglement properties remain same if one adds (subtracts) photons in mode \(1\) or in mode \(3\). Similarly for mode \(2\) or \(4\). We show here that the introduction of measurement breaks the symmetry -- we get more entanglement when photon is added in the unmeasured  modes compared to the one when photons are added to the measured one. Specifically, when the measurement is performed in mode \(4\), we obtain \(\mathcal{LG}^{add} (m_2)  > \mathcal{LG}^{add} (m_4)\). Interestingly, such a symmetry-breaking does not occur in the photon-subtracted state. Therefore, we show that the photon-addition procedure can be used as a detection method to identify the mode in which photons are added.
    
    \item With the increase of squeezing parameter, LGME increases as one expects. Interestingly, the gap between the LGME obtained in different modes decreases  with the moderate number of photons added in a single mode and with a moderate squeezing strength.  
    
    \end{enumerate}

\begin{figure}[t]
\includegraphics[width=\linewidth]{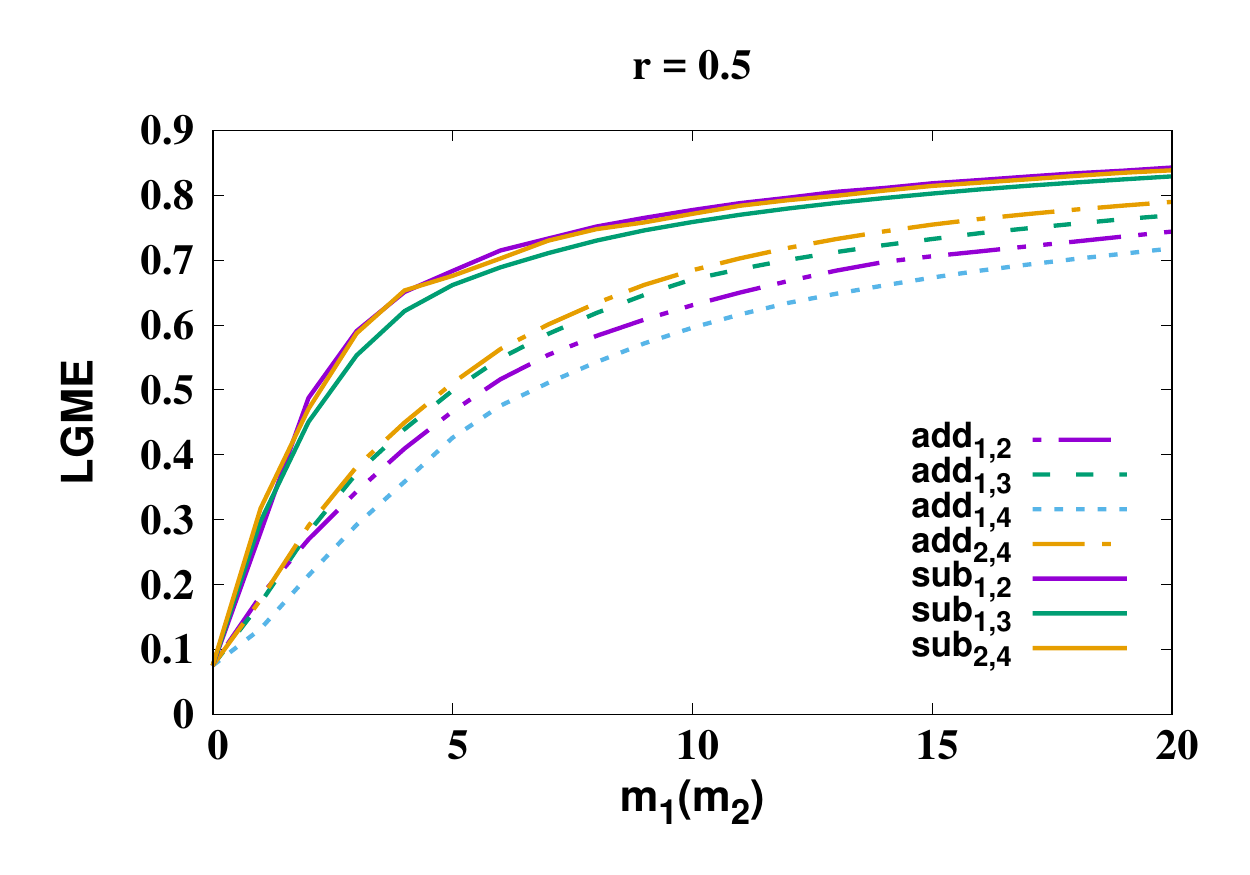}
\caption{ LGME  (y-axis) with respect to the equal number of photons added and subtracted in two different modes, i.e., \(m_i = m_j\). And correspondingly the legends are represented as add\(_{i,j}\) and sub\(_{i,j}\).   All other specifications are same as in Fig. \ref{fig:LGMEconatraint}.  }
\label{fig:LGMEequal}
\end{figure}

\noindent \emph{Two-mode operations.} Let us now move to the scenario when a fixed number of photons are added (subtracted) to (from) two modes, i.e. in \((m_i, m_j)\)-pair.  Suppose \(m_i + m_j = n\) is added  (subtracted) with \(i \neq j\). Here we can divide the situation into two categories as depicted in Figs. \ref{fig:LGMEconatraint} and \ref{fig:LGMEequal} - (1) measurement is performed neither on \(i\)th nor on the \(j\)th mode; (2) measurement is performed on, say \(j\)th mode. In our case \(j=4\). Like  single-mode operations, we again confirm that twin operations,  measuring and adding photons, in the same mode has adverse effects in localization of multimode entanglement.    

\emph{Case 1.} Let us first consider the pair \((m_1,m_3)\).  LGME monotonically increases till the point with \(m_i = m_j = n/2\), then decreases and  reaches to the same value since LGME coincides when addition happens either in mode \(1\) or \(3\). For mode \((1,2)\)-pair,   LGME clearly decreases with \(m_1\), as can be seen from Fig. \ref{fig:LGMEaddsubonemode}.  From Theorem 1, it is clear that LGME remains constant with \((m_i, m_{i+2})\)-pair for photon-subtracted state. On the other hand, in case of mode \((1,2)\)-pair, we observe that for \(n=20\), the maximum amount of GME can be localized when eight photons are subtracted from the first mode which reflects the fact that there is no symmetry in the exchange of \(m_1\) and \(m_2\).

\emph{Case 2.} In this picture, one of the parties, in our case, the fourth party performs the photon counting measurement.  Hence we are interested in the pair, \((m_i, m_4)\). In case of addition, we observe that  in case of \((m_2,m_4)\)-duo,  since the symmetry is broken due to the measurement on the fourth mode, localized entanglement of photon-added state at \((m_2=0,m_4=20)\)  is different than that of \((m_2=20,m_4=0\)). Moreover, from Fig. \ref{fig:LGMEaddsubonemode}, we find that entanglement is less when  photons are added in the measured mode compared to that of the unmeasured mode.  Hence we have \(\mathcal{LG} (m_2= 20, m_4= 0) > \mathcal{LG}(m_2 =0, m_4= 20)\) and it is an increasing curve with \(m_2\) although not monotonic. On the other hand, photon-subtraction does not break the symmetry of the initial state and so localized entanglement remains constant with \(m_2\) or \(m_4\) as shown in Fig. \ref{fig:LGMEconatraint}. 

\emph{Special Case.} We now consider the situation where equal number of photons are added or subtracted from modes. Note that it  is  a special configuration of the above mentioned case  since the photonic operations are equally distributed in the two modes. We highlight this because it neatly captures the LGME enhancement properties of bimodal photonic operations, see Fig. \ref{fig:LGMEequal} and again establishes that photon-subtraction is better than that of photon-addition, i.e. \(\mathcal{LG}^{sub} \geq \mathcal{LG}^{add}\).

\section{Conclusion}
\label{sec:con}

Characterization and manipulation of quantum resources form  integral parts of research in quantum information science facilitating optimal usage of the same. In this work, we focus  on concentrating genuine multimode entanglement (GME) in a multimode state from a state with higher number of modes using local projective measurements.  We were interested to quantify the amount of localizable GME (LGME) both in the Gaussian and non-Gaussian paradigms.  

We started off our analysis with a four-mode Gaussian  squeezed vacuum (FMSV) state and found the GME content that can be localized in the remaining three modes by employing optimal Gaussian measurements in the fourth mode. We showed that the maximum localization is possible when Gaussian measurement reduces to homodyne detection in either $x$ or $p$ quadrature and can be easily implemented experimentally. We also reported that if one  adopts a non-Gaussian measurement strategy like photon counting measurement scheme, it can lead to a higher  LGME in comparison to the optimal Gaussian measurement setting.

With a clear indication that non-Gaussianity aids in more effective localization of GME, we considered the  scenario where we computed LGME of non-Gaussian states by applying non-Gaussian measurements. The measurement strategy is kept fixed to be the photon counting operations as before, and the states under investigation are obtained by deGaussifying the FMSV state by performing photon addition and subtraction  in its various modes. These dual dose of non-Gaussianity enables localization of substantial GME from the optimal Gaussian case.  In addition to the overall enhancement of LGME, we also observed that for obtaining LGME,  local measurements can break the inherent symmetry, present in the original state in case of  addition while such a symmetry remains unaffected in the photon-subtracted state even after the measurement. It also revealed  that LGME of the photon-added states can be applied to identify the modes in which photons are added and  works for a moderate amount of squeezing in the original state,  achievable in current experiments. On the other hand, due to this property in the photon-subtracted state, there exist a situation when LGME remains constant with a fixed number of subtraction of photons from two modes. Our analysis indicates that the local measurements can uncover certain characteristics of states which are otherwise impossible to highlight. 


\acknowledgements

TD  acknowledges The ’International Centre for Theory of Quantum Technologies’ project (contract no. 2018/MAB/5). We acknowledge the support from the Interdisciplinary Cyber Physical Systems (ICPS) program of the Department of Science and Technology (DST), India, Grant No.: DST/ICPS/QuST/Theme- 1/2019/23.
%
%
%

\bibliography{bib}

%
%
%
\end{document}